# From Quantum Hall Effect To Topological Insulators

R. Tao

Department of Physics

Temple University, Philadelphia, PA 19122

The integral and fractional quantum Hall effects are among the most important discoveries in condensed matter physics in 1980s. The main results can be summarized in the conductance matrix, $\{\sigma_{ij}\} = \begin{pmatrix} \sigma_L & \sigma_H \\ -\sigma_H & \sigma_L \end{pmatrix}$. When the filling factor $\nu$ is an integer or some fractional value, such as 1/3, 2/5, 3/7, 2/3, 5/2,… , the conductance $\sigma_{ij}$ is quantized: $\sigma_L = 0$ and $\sigma_H = \nu e^2/h$. It is also proved that $\nu$ is related to the first Chern number, a topological invariant.

The novel properties of the quantum Hall system raised an important question: Can we have such important quantum properties in other systems? Especially, can we find a quantum system without magnetic field but maintaining the same properties in its conductance matrix? This is really the starting point for the topological insulators, too.

In fact, this issue was first discussed in our paper, published in Physical Review B, 1987. Here, I re-post this paper and want to emphasize that the physics discussed in our paper is general and profound. The essence of quantum Hall effect can be generalized to many quantum systems, either fermion system or boson system. If the quantum system has discrete energy spectrum and depends on two or more periodic parameters, the response coefficient will be the same as the conductance of quantum Hall effect, dissipation free and quantized.

# Integral and fractional quantization of a class of quantum systems

R. Tao and A. Widom

*Department of Physics, Northeastern University, Boston, Massachusetts 02115*

Abstract

The essence of integral and fractional quantization is studied. If a quantum system has a discrete energy spectrum and depends on several periodic variables, e.g. $\alpha_i$ and $\alpha_j$, the response coefficient $\eta_{ij}$ at zero temperature is dissipation free and either integrally quantized, when the ground state is non-degenerate, or fractionally quantized with a denominator $q$, when the ground state is $q$-fold degenerate. This quantization is a holonomy, the first Chern class. If the energy spectrum is continuous, $\eta_{ij}$ is dissipative and no longer quantized, but $\eta_{ji} - \eta_{ij}$ still is a quantized topological invariant. A singly connected Josephson junction coupled with a superconducting quantum-interference device ring is discussed as an example of the theory.

The integral[1] and fractional[2] quantum Hall effects are among the most important discoveries in condensed-matter physics in recent years. The high precision of the quantized values of the Hall conductance is so remarkable that a question is naturally raised: Are these integral and fractional quantization unique to the Hall system? Can other quantum systems also have such quantization? Under what condition do they hold true? The present paper is devoted to exploring the essence of the integral and fractional quantization. We have found that a class of quantum systems, either Fermi or Bose systems, is capable of having integral or fractional quantization which is a topological invariant. As an example of an application of the present theory, in the Appendix we show that a boson system, consisting of a singly connected Josephson junction coupled with a superconducting quantum-inference device (SQUID) ring, will have integral quantization.

Let us consider a quantum system whose Hamiltonian, $H(\alpha)$ contains a set of variables, $\vec{\alpha} = (\alpha_1,...,\alpha_n)$ $(n \geq 2)$. The force acting on the system by a change of variable $\alpha_i$ is given by $P_i = \partial H / \partial \alpha_i$. The flows are defined by the time rate of change of these variables, $\dot{\alpha}_i$. When $\dot{\alpha}_i$ varies, the system responds with forces $\delta \vec{P}$. The response coefficient $\eta_{ij}$ is defined by

$$\delta P_j = \sum_i \eta_{ji} \delta \dot{\alpha}_i . \qquad (1)$$

The kinetic coefficient $L_{ji}$ is the inverse of $\eta_{ij}$ and satisfies

$$\delta \dot{\alpha}_j = \sum_i L_{ji} \delta P_i . \qquad (2)$$

We will discuss both the static case in which $\delta \dot{\alpha}_i$ $(i = 1,...,n)$ are constants or varying very slowly and the frequency-dependent case in which $\delta \dot{\alpha}_i \sim e^{i\omega t}$. It must be noted that in some cases, for a normalized eigenstate of $H$, $\Psi_s$, $<\Psi_s | \partial H / \partial \alpha_i | \Psi_s >$ does not vanish. As a result the system has a net force even if $\delta\dot{\vec{\alpha}} = 0$. Since we are only interested in the forces induced by $\delta\dot{\vec{\alpha}}$, in such a case we must use

$$P_i = \frac{\partial H}{\partial \alpha_i} - \sum_s |\Psi_s\rangle\langle\Psi_s|\frac{\partial H}{\partial \alpha_i}|\Psi_s\rangle\langle\Psi_s|. \qquad (3)$$

For every point in the $\alpha$ space $(\alpha_1,...,\alpha_n)$, there is a Hamiltonian $H(\alpha)$ which characterizes the quantum system. Generally, at two different points in $\alpha$ space, the system is in different physical states. We are interested in a periodic case in which $\alpha_i$ ($i=1,...,n$) has a period $T_i$. At any two points, $(\alpha_1,...,\alpha_i,...)$ and $(\alpha_1,...,\alpha_i+T_i,...)$, the quantum system is in the same physical state. Then the whole $\alpha$ space can be represented by a unit cell $0 \leq \alpha_i \leq T_i$ ($i = 1, ... , n$). Any meaningful physical quantity, which is $\alpha$ independent, should be the average value on this unit cell. There are many quantum systems which have a periodic parameter space. For example, in the quantum Hall effect,[3-8] two solenoid fluxes $\phi_1$ and $\phi_2$ are periodic parameters, a result of gauge invariance. In the problem of charge transport discussed by Thouless *et al.*,[9,10] the Bloch wave number $k$ and time $t$ are two periodic variables. Some Bose systems, such as Josephson junctions in superconductors, also obviously possess such periodic parameters.

We assume that $\alpha_i$ has a small variation. This is a dynamic motion and may be represented by an additional term $H' = (\partial H/\partial \alpha_i)\delta\alpha_i = P_i\delta\alpha_i$, where $\delta\alpha_i$ is the changing part of $\alpha_i$. We introduce $\Pi_i = \int^t P_i dt$. Since the Hamiltonian

$$H_t = H - \Pi_i\delta\dot{\alpha}_i \qquad (4)$$

differs from $H + P_i \delta \alpha_i$ only by a full time derivative, $d(\Pi_i \delta \alpha_i)/dt$ they are equivalent.

According to linear-response theory,[11] the change of $P_j$, $\delta P_j$, arising from $\delta \dot{\alpha}_i$ is given by

$$\delta P_j(t) = \frac{1}{i\hbar} \int_{-\infty}^{t} Tr[\rho, \Pi_i] P_j(t-\tau) \delta \dot{\alpha}_i(\tau) d\tau, \tag{5}$$

where $P_j(t) = e^{iHt/\hbar} P_j e^{-iHt/\hbar}$. Let the density operator $\rho$ be canonical, $\rho = e^{-\beta H}/Z$ where $\beta = 1/k_B T$ and $Z = Tr(e^{-\beta H})$. We then have

$$[\rho, \Pi_i] = \int_0^\beta \rho e^{\lambda H} [\Pi_i, H] e^{-\lambda H} d\lambda = i\hbar \int_0^\beta \rho \dot{\Pi}_i(-i\hbar\lambda) d\lambda = i\hbar \int_0^\beta \rho P_i(-i\hbar\lambda) d\lambda. \tag{6}$$

Equation (5) now can be written as

$$\delta P_j(t) = \int_{-\infty}^{t} \phi_{ji}(t-\tau) \delta \dot{\alpha}_i(\tau) d\tau, \tag{7}$$

where the response function is given by

$$\phi_{ji}(t) = \int_0^\beta d\lambda Tr[\rho P_i(-i\hbar\lambda) P_j(t)]. \tag{8}$$

The response coefficient at $\delta \dot{\alpha}_i \sim e^{i\omega t}$ is the Fourier transform of $\phi_{ji}$,

$$\eta_{ji}(\omega) = \int_0^\infty e^{-i\omega t} dt \int_0^\beta d\lambda Tr[\rho P_i(-i\hbar\lambda) P_j(t)]. \tag{9}$$

In the static case, the response coefficient is given by

$$\eta_{ji} = \int_0^\infty dt \int_0^\beta d\lambda Tr[\rho P_i(-i\hbar\lambda) P_j(t)]. \tag{10}$$

Using the set of eigenfunctions of $H(\alpha)$, $\{\Psi_s\}$ and Eq.(3), we expand Eq.(10),

$$\eta_{ji} = \frac{1}{Z} \sum_{\substack{s,l \\ (s \neq l)}} \int_0^\infty dt \int_0^\beta d\lambda e^{-\beta E_s + (\lambda - it/\hbar)(E_s - E_l)} \left\langle \Psi_s \left| \frac{\partial H}{\partial \alpha_i} \right| \Psi_l \right\rangle \left\langle \Psi_l \left| \frac{\partial H}{\partial \alpha_j} \right| \Psi_s \right\rangle \tag{11}$$

The formula $\int_0^\infty dt e^{ikt} = i/k + \pi \delta(k)$ enables us to simplify Eq. (11) further. We divide $\eta_{ji}$ into two parts: an antisymmetric part $\eta_{ji}^a = -\eta_{ij}^a$ and symmetric part $\eta_{ji}^s = \eta_{ij}^s$, given by

$$\eta_{ji}^a = \frac{i\hbar}{Z} \sum_s e^{-\beta E_s} \left[ \left\langle \frac{\partial \Psi_s}{\partial \alpha_i} \middle| \frac{\partial \Psi_s}{\partial \alpha_j} \right\rangle - \left\langle \frac{\partial \Psi_s}{\partial \alpha_j} \middle| \frac{\partial \Psi_s}{\partial \alpha_i} \right\rangle \right], \tag{12}$$

$$\eta_{ji}^{s} = \frac{\hbar\pi\beta}{Z}\sum_{\substack{s,l\\(s\neq l)}} e^{-\beta E_s}\delta(E_s - E_l)\left\langle\Psi_s|\frac{\partial H}{\partial\alpha_i}|\Psi_l\right\rangle\left\langle\Psi_l|\frac{\partial H}{\partial\alpha_j}|\Psi_s\right\rangle. \tag{13}$$

It is easy to verify that $\eta_{ji} = \eta_{ji}^{a} + \eta_{ji}^{s}$ satisfies the Onsager relation.[11] The symmetric part $\eta_{ji}^{s}$ gives rise to a dissipation. Generally, the static $\eta_{ji}^{a}$ and $\eta_{ji}^{s}$ are functions of $\alpha$. But if the $\alpha$ space is periodic and represented by a unit cell $0 \leq \alpha_i \leq T_i, (i=1,...,n)$ the meaningful $\eta_{ji}$ should be the average value on the unit cell; then we write

$$\eta_{ji}^{a} = \frac{i\hbar}{ZT_iT_j}\sum_s e^{-\beta E_s}\int_0^{T_i} d\alpha_i \int_0^{T_j} d\alpha_j \left[\left\langle\frac{\partial\Psi_s}{\partial\alpha_i}\Big|\frac{\partial\Psi_s}{\partial\alpha_j}\right\rangle - \left\langle\frac{\partial\Psi_s}{\partial\alpha_j}\Big|\frac{\partial\Psi_s}{\partial\alpha_i}\right\rangle\right] \tag{14}$$

The same manipulation enables us to write the average $\eta_{ji}(\omega)$ in Eq. (9) into

$$\eta_{ji}(\omega) = \frac{\hbar}{ZT_iT_j}\int_0^{T_i} d\alpha_i \int_0^{T_j} d\alpha_j [i\sum_n e^{-\beta E_n}\left[\left\langle\frac{\partial\Psi_n}{\partial\alpha_i}\Big|\frac{H-E_n}{H-E_n-\hbar\omega}\Big|\frac{\partial\Psi_n}{\partial\alpha_j}\right\rangle - \left\langle\frac{\partial\Psi_s}{\partial\alpha_j}\Big|\frac{H-E_n}{H-E_n+\hbar\omega}\Big|\frac{\partial\Psi_s}{\partial\alpha_i}\right\rangle\right]$$

$$+\sum_{\substack{s,l\\(s\neq l)}} \pi\delta(E_s - E_l + \hbar\omega)(e^{-\beta E_l} - e^{-\beta E_s})(E_s - E_l)\left\langle\frac{\partial\Psi_s}{\partial\alpha_i}|\Psi_l\right\rangle\left\langle\Psi_l|\frac{\partial\Psi_s}{\partial\alpha_j}\right\rangle$$

$$(15)$$

At zero temperature, $\beta \to \infty$, only the ground state has a contribution in Eq. (14). If the ground state is non-degenerate, we have $\eta_{ji}^{a}$ integrally quantized,

$$\eta_{ji}^{a} = 2\pi\hbar(\gamma_{ji})/T_iT_j. \tag{16}$$

where the integer $\gamma_{ji}$ is given by

$$\frac{i}{2\pi}\int_0^{T_i}\int_0^{T_j}\left[\left\langle\frac{\partial\Psi_0}{\partial\alpha_j}\Big|\frac{\partial\Psi_0}{\partial\alpha_i}\right\rangle - \left\langle\frac{\partial\Psi_0}{\partial\alpha_i}\Big|\frac{\partial\Psi_0}{\partial\alpha_j}\right\rangle\right]d\alpha_i d\alpha_j = \frac{i}{2\pi}\oint\left\langle\Psi_0|\frac{\partial\Psi_0}{\partial\vec{\alpha}}\right\rangle d\vec{\alpha}. \tag{17}$$

Stoke's theorem is used here. The last integration is along the contour of the unit cell $0 \leq \alpha_i \leq T_i$ and $0 \leq \alpha_j \leq T_j$. $\Psi_0$ is the normalized non-degenerate ground state. This integer $\gamma_{ji}$ is a holonomy. As one moves along the contour, $\Psi_0$ yields a line of bundle over the parameter space, which has a natural Hermitian connection, studied by Bott, Chern,[12] and Simon.[5] Then $\gamma_{ji}$ is the conventional integral of curvature which is just the Chern class of the connection. Because the unit cell is a torus in this problem, the first Chern class is an integer.

If the ground state is $q$-fold degenerate, where $q$ is a finite integer, $\eta_{ji}^a$ in Eq. (14) at zero temperature is fractionally quantized with denominator $q$,

$$\eta_{ji}^a = 2\pi\hbar(\gamma_{ji}^{'}/q)/T_iT_j \qquad (18)$$

where the integer $\gamma_{ji}^{'}$ is now given by

$$\frac{i}{2\pi}\sum_{k=1}^{q}\int_0^{T_i}\int_0^{T_j}\left[\left\langle\frac{\partial\Psi_0^{(k)}}{\partial\alpha_j}\bigg|\frac{\partial\Psi_0^{(k)}}{\partial\alpha_i}\right\rangle - \left\langle\frac{\partial\Psi_0^{(k)}}{\partial\alpha_i}\bigg|\frac{\partial\Psi_0^{(k)}}{\partial\alpha_j}\right\rangle\right]d\alpha_i d\alpha_j. \qquad (19)$$

$\Psi_0^{(k)}$ ($k=1,...,q$) are $q$ orthonormal degenerate ground states. The appearance of $q$ in the denominator comes from the partition function $Z$ in Eq.(14). The integer $\gamma_{ji}^{'}$ is also the first Chern class. It should be noted that since our discussion is independent of the statistics of the quantum system, the above conclusions must be true for both Fermi systems and Bose systems.

Suppose that the quantum system has a discrete energy spectrum, which is especially true when the system is finite. Then there is an energy gap $\Delta$ separating $\Psi_0$ from excited states. At zero temperature, from Eq.(13), $\eta_{ji}^s = 0$. Thus, under such a condition, $\eta_{ji} = \eta_{ji}^a$ which is represented by Eq. (16) or (18), integrally or fractionally quantized, and dissipation free. From Eq. (14), at a very low temperature, the contribution from excited states has the order of $e^{-\Delta/k_BT}$. If $k_BT < \Delta$, this correction is negligible. As a result, at a very low temperature, $\eta_{ji}$ still has an accurate integral or fractional quantization.

When the energy spectrum of the system is continuous, $\eta_{ji}^s$ is not vanishing, so $\eta_{ji}$ has dissipation and itself is no longer quantized. But at zero temperature, from the above discussion,

$$\eta_{ji} - \eta_{ij} = 2\eta_{ji}^a \qquad (20)$$

is always integrally or fractionally quantized, regardless of the situation of the energy gap.

At a finite temperature, the integrals in Eq. (14) still give integers if $\Psi_s$ is non-degenerate, and fractions if $\Psi_s$ is degenerate, but $\eta_{ji}^a$ is the thermal average of these integers and fractions; therefore, it itself is no longer integrally or fractionally quantized. From Eq. (15), it is easy to know that at a high frequency $\eta_{ji}(\omega)$ is not quantized, either.

## ACKNOWLEDGMENT

R.T.'s work was supported in part by a grant from Northeastern University's Research and Scholarship Fund.

## APPENDIX

As an example of application of the present theory, we consider a singly connected Josephson junction coupled with a SQUID ring. The system, shown in Fig. 1, has two periodic parameters, $\theta_1$ and $\theta_2$, which are the phase differences across the two Josephson junctions. At low temperature, there is an energy gap separating the superconducting state from the normal state. It is noted that this is a boson system, completely different from the quantum-Hall electron system. Now we inject a current, $J_2$, into the system to produce a small voltage $V_2$ and study the relationship between $V_2$ and the induced current $J_1$ in the SQUID. We have

$$J_1 = \eta_{12} V_2. \qquad (A1)$$

Since current $J_1$ is given by[14]

$$J_1 = -\frac{2e}{\hbar}\left[\frac{\partial H}{\partial \theta_1}\right], \qquad (A2)$$

Where $H$ the Hamiltonian of the system and voltage is $V_2$ is given by

$$V_2 = -\frac{\hbar \dot{\theta}_2}{2e}, \qquad (A3)$$

Then according to our above general theory, $\eta_{12}$ will be found integrally quantized.

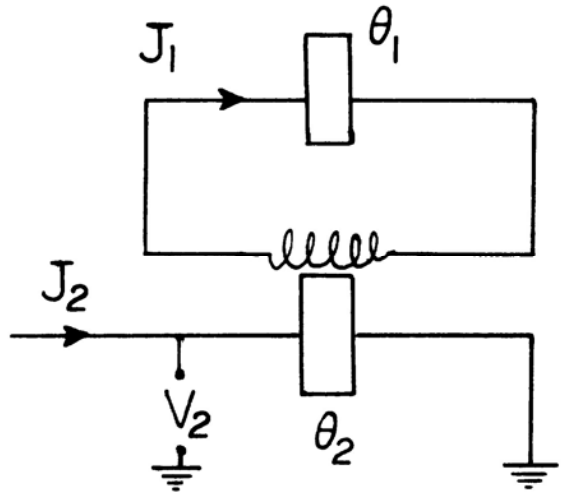

Fig.1. a single connected Josephson junction coupled with a SQUID ring.

**References**

1. K. von Klitzing, G. Dora, and M. Pepper, Phys. Rev. Lett. **45**, 494 (1980); R. B. Laughlin, Phys. Rev. **23**, 5632 (1981); B. I. Halperin, *ibid.* 25, 2185 (1982).

2. D. C. Tsui, H. L. Stormer, and A. C. Gossard, Phys. Rev. Lett.**48**, 1559 (1982); R. B. Laughlin, ibid. **50**, 1395 (1983); F. D. M. Haldane, *ibid.* **51**, 605 (1983); B. I. Halperin, *ibid.* **52**, 1583 (1984); S. Kivelsion, C. Kallin, D. P. Aravos, and J. R. Schrieffer, ibid.**56**, 873 (1986); R. Tao and Y. S. Wu, Phys. Rev. B **30**, 1097 (1984).